\documentclass[aps, pra, floatfix, pdftex, showpacs, 10pt]{revtex4-1}
\usepackage{eurosym}
\usepackage{dcolumn}
\usepackage{bm}
\usepackage{amssymb, amsmath}
\usepackage{graphicx}
\usepackage{hyperref}
\usepackage{epstopdf}
\usepackage{subfigure}
\usepackage{color}
\usepackage[normalem]{ulem}
\usepackage[utf8]{inputenc}
\usepackage[sort&compress]{natbib}

\pdfoutput=1

\begin{document}

\title{Symmetry breakings in dual-core systems with double-spot localization of nonlinearity}

\author{Krzysztof B. Zegadlo$^{1}$*, Nguyen Viet Hung
$^{2}$, Aleksandr Ramaniuk $^{3}$, Marek Trippenbach $^{3}$ and Boris A. Malomed $^{4}$}

\affiliation{%
$^{1}$ Faculty of Physics, Astronomy and Applied Computer Science, Jagiellonian University, Lojasiewicza 11, PL-30-348, Krakow, Poland;\\
$^{2}$ Advanced Institute for Science and Technology, Hanoi University of Science and Technology, Hanoi, Vietnam;\\
$^{3}$ Faculty of Physics, University of Warsaw, ul. Pasteura 5, PL-02-093, Warszawa, Poland;\\
$^{4}$ Department of Physical Electronics, School of Electrical Engineering, Faculty of Engineering, Tel Aviv University, Tel Aviv 69978, Israel}

\begin{abstract}
We introduce a dual-core system with double symmetry, one between the cores,
and one along each core, imposed by the spatial modulation of local
nonlinearity in the form of two tightly localized spots, which may be
approximated by a pair of ideal delta-functions. The analysis aims to
investigate effects of spontaneous symmetry breaking in such systems.
Stationary one-dimensional modes are constructed in an implicit analytical
form. These solutions include symmetric ones, as well as modes with
spontaneously broken inter-core and along-the-cores symmetries. Solutions
featuring the simultaneous (double) breaking of both symmetries are produced
too. In the model with the ideal delta-functions, all species of the
asymmetric modes are found to be unstable. However, numerical consideration
of a two-dimensional extension of the system, which includes symmetric cores
with a nonzero transverse thickness, and the nonlinearity-localization spots
of a small finite size, produces stable asymmetric modes of all the types,
realizing the separate breaking of each symmetry, and states featuring
simultaneous (double) breaking of both symmetries.
\end{abstract}

\maketitle

\section{Introduction}

Dual-core couplers represent one of basic types of optical waveguides \cite%
{Huang}. In most cases, the couplers are realized as twin-core optical
fibers \cite{fiber-coupler,fiber-coupler2} (i.e., in the \textit{temporal
domain}), as well as in double-layer planar waveguides (see, e.g., Ref. \cite%
{planar}), i.e., in the \textit{spatial domain}. Similar to two-core fiber
waveguides for optical waves are dual-core cigar-shaped (strongly elongated)
traps for matter waves in atomic Bose-Einstein condensates (BECs) \cite%
{Randy}. Transmission of matter waves in the latter setting have been
studied theoretically \cite{BEC1,Luca2,Marek}, making use of the fact that
the Gross-Pitaevskii equation (GPE)\ for the mean-field wave function of the
matter waves in BEC \cite{BEC} is identical to the propagation equation for
the complex amplitude electromagnetic waves in optical waveguides.

If the intrinsic nonlinearity in the parallel-coupled cores is strong
enough, the field exchange between them is affected by the intensity of the
guided waves \cite{Jensen,Maier}. This effect may be used as a basis for the
design of switching devices \cite{switch1}-\cite{review} and other
applications. Further, a fundamental property of nonlinear couplers with
mutually symmetric cores is the \textit{symmetry-breaking bifurcation},
which destabilizes obvious symmetric modes in the dual-core system, and
gives rise to asymmetric states. The symmetry breaking was analyzed for
temporally uniform states (continuous waves) in dual-core nonlinear optical
fibers \cite{Snyder}, as well as for solitons in the same system \cite%
{Wabnitz}-\cite{Kin}. Some results obtained in this area were summarized
in early reviews \cite{Wabnitz2} and \cite{Progress}, and in a recent review
article \cite{in-Peng}.

The Kerr nonlinearity in the dual-core system gives rise to the
symmetry-breaking bifurcation for solitons of the \textit{subcritical} type,
with emerging unstable branches of asymmetric modes going backward (in the
direction of decreasing nonlinearity strength), and then turning forward
\cite{bifurcations}. The branches stabilize themselves at the turning
points. On the other hand, the bifurcation for continuous waves, in the same
system, is of the\textit{\ supercritical} type. It creates stable branches
of asymmetric states, which immediately go forward (in the direction of
strengthening nonlinearity).

The objective of the present paper is to report new result for the ordinary
and \emph{double} symmetry breakings in dual-core systems, in which the
additional symmetry in each core is introduced by assuming that its
intrinsic nonlinearity is subject to the spatial modulation, being
concentrated at two mutually symmetric spatially separated points. Such a
setting may be implemented in optics, by means of spatially inhomogeneous
distribution of a nonlinearity-enhancing dopant in the waveguiding cores
\cite{Kip}, as well as in BEC, using Feshbach resonance controlled by laser
beams focused at the nonlinearity-concentration spots \cite{Feshbach}. In
particular, as concerns this phenomenology in quantum gases, spontaneous
symmetry breaking between \emph{degenerate ground states} is an important
feature of quantum phase transitions in bosonic systems. The understanding
and control of symmetries of quantum systems may lead to realization of new
quantum phases, which are subjects of great interest to fundamental and
applied studies alike. It is relevant to mention that, while ordinary
spontaneous symmetry breaking was considered in various settings \cite{book}%
, effects of \emph{double symmetry breaking} were previously addressed only
in few cases, see, e.g., Ref. \cite{Raymond}.

The rest of the paper is organized as follows. The model is introduced in
Section 2, where its general analytical solution, available (in an implicit
form of coupled cubic equations for four constituent amplitudes, see Eq. (%
\ref{cubic equations}) below) in the case when the nonlinearity in each core
is represented by a symmetric pair of ideal delta-functions (as suggested by
Ref. \cite{Dong}, in terms of the single-core model), is given too. Novel
results are reported in Section 3, for states which maintain the spatial
symmetry between the two delta-functional spots in each core, while breaking
the inter-core symmetry via a subcritical bifurcation. In addition, a
simplified model with the nonlinearity-modulation profile represented by a
single delta-function is considered in Section 3 too. In Section 4, we
address the breaking of the spatial symmetry (between the two
nonlinearity-concentration spots) in the case when the symmetry between the
cores is kept unbroken, hence the system is reduced to the single-core one.
A similar setting was considered before in Ref. \cite{Dong}, therefore in
Section IV we briefly recapitulate results for this case.

While in all the versions of the model considered in Sections 3 and 4
broken-symmetry states are unstable, essentially new results are reported in
Section 5 for the two-dimensional version of the two-core system, with a
small but finite transverse thickness of the channels (cores), and a small
but finite width of the two nonlinearity spots. In this case, the system
supports \emph{stable} modes with broken inter-core symmetry, or the
intra-core one symmetry between the two nonlinearity spots, as well as
stable states which feature \emph{double }symmetry breaking.

The paper is concluded by Section 6.

\section{Model}

The purpose of our study is to investigate the interplay of inter- and
intra-channel symmetry breaking of soliton modes in a pair of
linearly-coupled planar nonlinear optical waveguides, as well as in a
Bose-Einstein condensate (BEC) loaded into a dual-cigar-shaped potential
trap, with two parallel cores coupled by tunneling of atoms. In either case,
the intra-channel symmetry breaking is considered with respect to a
symmetric pair of spots at which the local nonlinearity is concentrated. The
model of this setting is based on a system of linearly-coupled nonlinear Schr%
\"{o}dinger equations, alias Gross-Pitaevskii equations (GPEs), in terms of
the BEC realization \cite{BEC}. In particular, the coupled GPEs can be
derived, as usual, from the underlying 3D GPE by means of dimensional
reduction to the effective 1D form, under the action of tight transverse
confinement \cite{Luca,Delgado}. The result, written in the scaled form, is
the following system of equations for wave functions $\psi $ and $\phi $ in
the two tunnel-coupled cores:

\begin{eqnarray}
i\frac{\partial \phi }{\partial t} &=&-\frac{1}{2}\frac{\partial ^{2}\phi }{%
\partial x^{2}}+g\left( x\right) |\phi |^{2}\phi -\kappa \psi ,  \notag \\
i\frac{\partial \psi }{\partial t} &=&-\frac{1}{2}\frac{\partial ^{2}\psi }{%
\partial x^{2}}+g\left( x\right) |\psi |^{2}\psi -\kappa \phi ,
\label{NLSEs system}
\end{eqnarray}%
where $x$ is the longitudinal coordinate, $g(x)$ is the local nonlinearity
coefficient, and $\kappa $ is the coupling strength. In terms of a dual-core
planar optical waveguide, $\phi $ and $\psi $ are amplitudes of the
electromagnetic waves in two cores, $x$ is the transverse coordinate, while
time $t$ is replaced by the propagation distance ($z$). Equations (\ref%
{NLSEs system})\ conserves the total norm,
\begin{equation}
N=\int_{-\infty }^{+\infty }\left[ |\phi (x)|^{2}+|\psi (x)|^{2}\right] dx
\label{N}
\end{equation}%
(alias the total power, in terms of the optical waveguides) and the
Hamiltonian,
\begin{equation}
H=\frac{1}{2}\int_{-\infty }^{+\infty }\left[ |\phi _{x}|^{2}+|\psi
_{x}|^{2}+g\left( x\right) \left( |\phi |^{4}+|\psi |^{4}\right) -2\kappa
\left( \phi \psi ^{\ast }+\phi ^{\ast }\psi \right) \right] dx,  \label{H}
\end{equation}%
where the asterisk stands for the complex conjugate wave function.

As said above, our aim is to consider the setting in which the self-focusing
nonlinearity is concentrated, in each core, at two tightly localized spots.
To the first approximation, following Refs. \cite{Dong,equal_potentials,Shnir,Han Pu}, this situation may be modelled by introducing the local nonlinearity coefficient
taken in the form of a sum of two delta-functions:
\begin{equation}
g\left( x\right) = - \left[\delta \left( x+1\right) +\delta \left( x-1\right)\right] ,
\label{delta}
\end{equation}%
with the separation between them fixed to be $2$ by means of the remaining
scaling invariance of Eq. (\ref{NLSEs system}). This model, as well as its
numerical implementation with $\delta $-functions replaced by the standard
approximation in the form of narrow Gaussians, makes it easier to analyze
the double symmetry breaking, between the two cores (represented by $\phi $
and $\psi $), and between vicinities of points $x=-1$ and $+1$.

Stationary solutions with real chemical potential $\mu $ ($-\mu $ is the
propagation constant in the model of the optical waveguide) are looked for
in the usual form,
\begin{equation}
\phi \left( x,t\right) =e^{-i\mu t}u\left( x\right) ,~\psi \left( x,t\right)
=e^{-i\mu t}v\left( x\right) ,  \label{mu}
\end{equation}%
and substitute it in Eqs. (\ref{NLSEs system}) to obtain a system of coupled
equations for real functions $u(x)$ and $v(x)$:
\begin{eqnarray}
\mu u &=&-\frac{1}{2}u^{\prime \prime }-\left[ \delta \left( x+1\right)
+\delta \left( x-1\right) \right] u^{3}-\kappa v,  \notag \\
\mu v &=&-\frac{1}{2}v^{\prime \prime }-\left[ \delta \left( x+1\right)
+\delta \left( x-1\right) \right] v^{3}-\kappa u.
\label{stationary NLSEs for deltas}
\end{eqnarray}%
Everywhere except for infinitesimal vicinities of points $x=\pm 1$, Eq. (\ref%
{stationary NLSEs for deltas}) is linear, and it may be diagonalized for
symmetric and antisymmetric solutions $w_{1}\equiv u+v$ and $w_{2}\equiv u-v$%
:
\begin{eqnarray}
-\mu _{+}w_{1}+w_{1}^{\prime \prime } &=&0  \notag \\
-\mu _{-}w_{2}+w_{2}^{\prime \prime } &=&0,  \label{uncoupled system}
\end{eqnarray}%
where $\mu _{\pm }\equiv -2(\mu \pm \kappa )$. Restricting the consideration
to localized modes, with $\mu _{\pm }>0$, and following the lines of Ref.
\cite{Dong}, where the analysis was developed for the single-component
model, one can write respective solutions to Eq. (\ref{uncoupled system}) as%
\begin{eqnarray}
u &=&\left\{
\begin{array}{c}
Be^{\sqrt{\mu _{+}}\left( x+1\right) }+De^{\sqrt{\mu _{-}}\left( x+1\right)
}, \\
\mathrm{at~}x<-1, \\
B_{0}e^{\sqrt{\mu _{+}}\left( x+1\right) }+D_{0}e^{\sqrt{\mu _{-}}\left(
x+1\right) }+A_{0}e^{-\sqrt{\mu _{+}}\left( x-1\right) }+C_{0}e^{-\sqrt{\mu
_{-}}\left( x-1\right) }, \\
\mathrm{at~}|x|~\leq 1, \\
Ae^{-\sqrt{\mu _{+}}\left( x-1\right) }+Ce^{-\sqrt{\mu _{-}}\left(
x-1\right) }, \\
\mathrm{at~}x>1.%
\end{array}%
\right.  \notag \\
v &=&\left\{
\begin{array}{c}
Be^{\sqrt{\mu _{+}}\left( x+1\right) }-De^{\sqrt{\mu _{-}}\left( x+1\right)
}, \\
\mathrm{at~}x<-1, \\
B_{0}e^{\sqrt{\mu _{+}}\left( x+1\right) }-D_{0}e^{\sqrt{\mu _{-}}\left(
x+1\right) }+A_{0}e^{-\sqrt{\mu _{+}}\left( x-1\right) }-C_{0}e^{-\sqrt{\mu
_{-}}\left( x-1\right) }, \\
\mathrm{at~}|x|~\leq 1, \\
Ae^{-\sqrt{\mu _{+}}\left( x-1\right) }-Ce^{-\sqrt{\mu _{-}}\left(
x-1\right) }, \\
\mathrm{at~}x>1.%
\end{array}%
\right.  \label{NLSEs solutions}
\end{eqnarray}

The effect of localized nonlinear terms is introduced through the continuity
and jump relations for the wave functions and their first derivatives,
respectively, at $x=\pm 1$. The former condition allows one to express
amplitudes $A_{0}$, $B_{0}$, $C_{0}$ and $D_{0}$ in terms of $A$, $B$, $C$
and $D$, in Eq. (\ref{NLSEs solutions}):%
\begin{eqnarray}
A_{0} &=&\frac{e^{2\sqrt{\mu _{+}}}B-A}{e^{4\sqrt{\mu _{+}}}-1},B_{0}=\frac{%
e^{2\sqrt{\mu _{+}}}A-B}{e^{4\sqrt{\mu _{+}}}-1},  \notag \\
C_{0} &=&\frac{e^{2\sqrt{\mu _{-}}}D-C}{e^{4\sqrt{\mu _{-}}}-1},D_{0}=\frac{%
e^{2\sqrt{\mu _{-}}}C-D}{e^{4\sqrt{\mu _{-}}}-1}.
\label{elimination of 4 amplitudes}
\end{eqnarray}%
Expressions for the jumps of the derivatives, $\Delta \left( u^{\prime
},v^{\prime }\right) |_{x=\pm 1}$, are obtained by integration of Eq. (\ref%
{stationary NLSEs for deltas}) in infinitesimal vicinities of $x=\pm 1$:
\begin{equation}
\Delta \left( u^{\prime }\right) |_{x=\pm 1}=-2\left( u|_{x=\pm 1}\right)
^{3},~\Delta \left( v^{\prime }\right) |_{x=\pm 1}=-2\left( v|_{x=\pm
1}\right) ^{3}.  \label{Delta}
\end{equation}

The substitution of solutions (\ref{NLSEs solutions}) in relations (\ref%
{Delta})\ generates four cubic equations for amplitudes:

\begin{eqnarray}
\sqrt{\mu_{+}}e^{2\sqrt{\mu_{+}}}\left(e^{2\sqrt{\mu_{+}}}B-A\right)=%
\left(e^{4\sqrt{\mu_{+}}}-1\right)B\left(B^{2}+3D^{2}\right),  \notag \\
\sqrt{\mu_{+}}e^{2\sqrt{\mu_{+}}}\left(e^{2\sqrt{\mu_{+}}}A-B\right)=%
\left(e^{4\sqrt{\mu_{+}}}-1\right)A\left(A^{2}+3C^{2}\right),  \notag \\
\sqrt{\mu_{-}}e^{2\sqrt{\mu_{-}}}\left(e^{2\sqrt{\mu_{-}}}D-C\right)=%
\left(e^{4\sqrt{\mu_{-}}}-1\right)D\left(D^{2}+3B^{2}\right),  \notag \\
\sqrt{\mu_{-}}e^{2\sqrt{\mu_{-}}}\left(e^{2\sqrt{\mu_{-}}}C-D\right)=%
\left(e^{4\sqrt{\mu_{-}}}-1\right)C\left(C^{2}+3A^{2}\right).
\label{cubic equations}
\end{eqnarray}

\begin{figure}[th]
\centering
\includegraphics[scale=0.35]{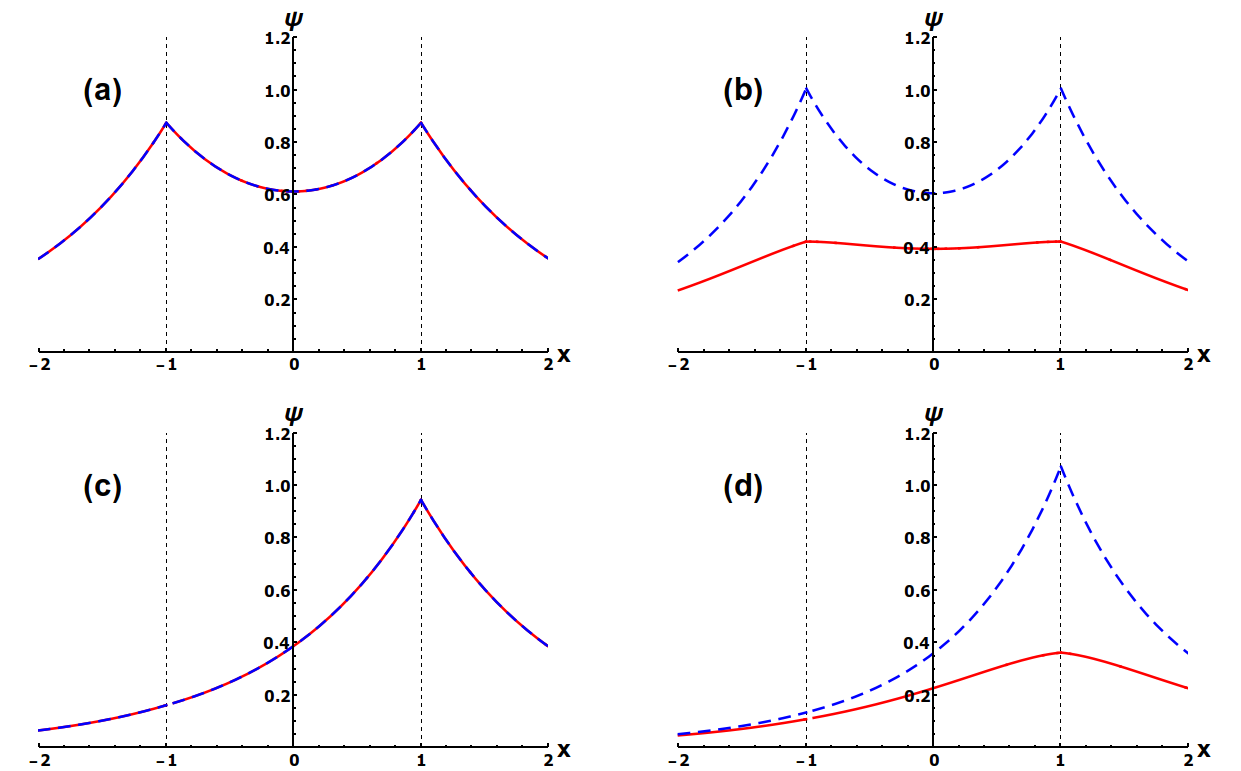}
\caption{Possible solutions of \eqref{cubic equations} plotted as the wavefunctions $\psi$ (red, solid) as well as $\phi$ (blue, dashed) corresponding to the cases when: (a) both symmetries are conserved, (b) the symmetry between both cores is broken, (c) only intra-core-symmetry is broken and (d) both symmetries are broken. These wavefunctions are numerical solutions of \eqref{cubic equations} for $\mu=-0.9$ and $\kappa=0.5$}
\label{4_solutions}
\end{figure}

\begin{figure}[th]
\centering
\includegraphics[scale=0.35]{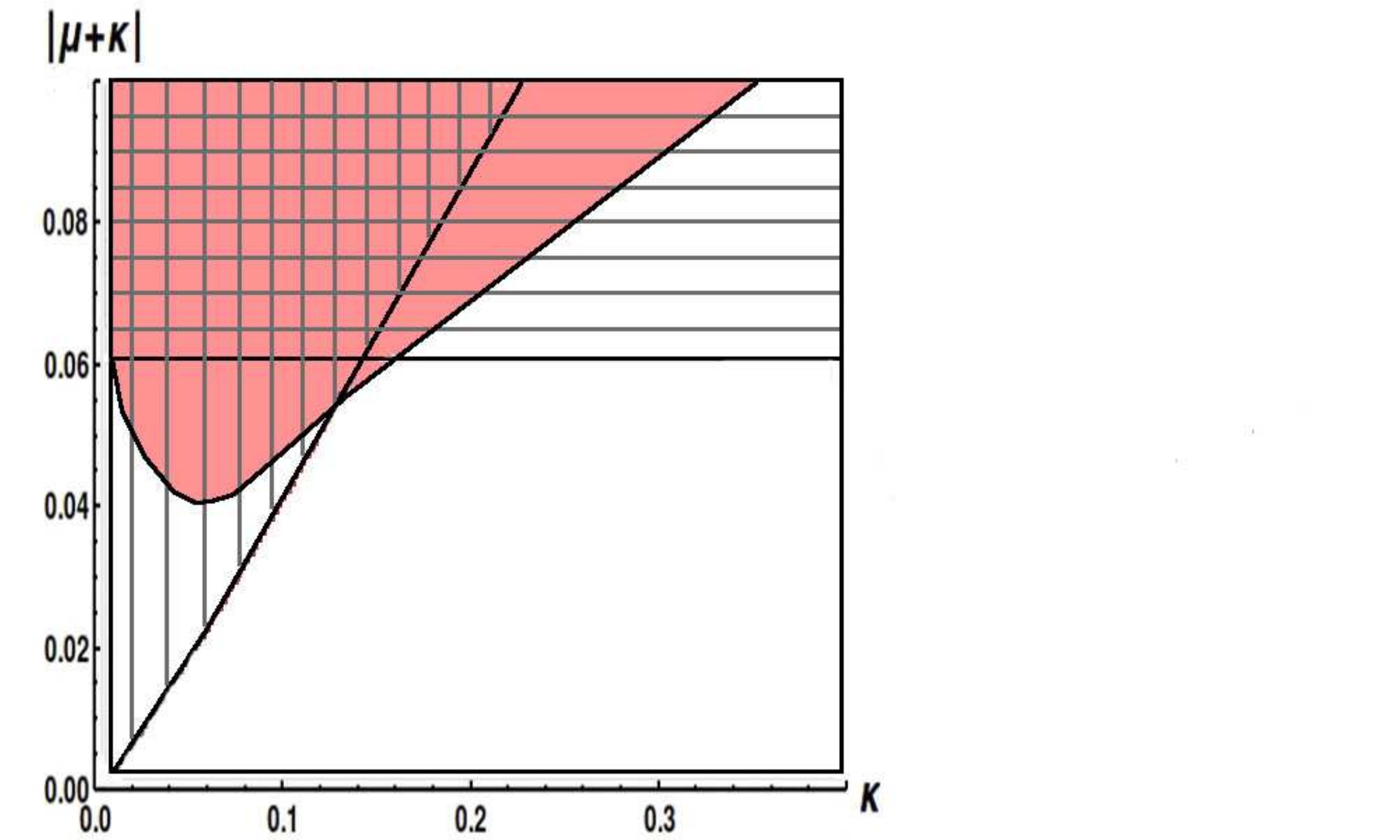}
\caption{The phase diagram, in the $\left( \protect\kappa ,\left\vert
\protect\mu +\protect\kappa \right\vert \right) $ plane, illustrating
regions where solutions of different types exist.
Solutions with unbroken symmetry exist everywhere. Above horizontal line $%
\protect\mu _{+}=0.06$, in the horizontally shaded region, one additionally finds solutions
with broken intra-channel symmetry. In the vertically shaded region, there appear
 solutions with broken symmetry between the channels ($u$ and $v$). In the pink
region, there exists solutions both symmetries simultaneously broken . Note that there is a
small triangular region where only fully symmetric solutions and ones
with both symmetries broken can be found (the unshaded pink area).}
\label{freshmap}
\end{figure}

The solutions of the system of nonlinear equations (\ref{stationary NLSEs
for deltas}), in the form defined by Eqs. (\ref{elimination of 4 amplitudes}%
) and (\ref{cubic equations}), are the main analytical result of the present
work. They are analyzed through the rest of the paper in the context
of the twofold spontaneous symmetry breaking in the system's ground state,
between the areas of $x>0$ and $x<0$, and between components $u$ and $v$ .
We show that, depending on parameters of the system, $\kappa $ and $\mu $,
solutions of different kinds (co)exist, {\it viz}., completely symmetric ones,
or solutions which break the symmetry between the channels, or break
the parity (symmetry with respect to the origin) in each channel, or, finally,
solutions breaking both symmetries (cf. functions presented in Fig. \ref{4_solutions}).

Coupled algebraic equations (\ref{cubic equations}), to which the use of the
ideal delta-functions makes it possible to reduce the solution of Eq. (\ref%
{stationary NLSEs for deltas}), cannot be solved analytically (unlike the
explicitly solution which is possible in the single-component model \cite%
{Dong}). It is, nevertheless, relatively easy to find all physically
relevant solutions of Eq.~(\ref{cubic equations}) numerically, and thus
classify all the solutions according to their symmetry.

In general, there are $3^4=81$ solutions, but many of them are highly
degenerate. We summarize the results in the schematic form in Fig.~(\ref%
{freshmap}), in the parameter plane of linear coupling $\kappa $ and shifted
chemical potential $|\mu +\kappa |$. Ground-state solutions with both
symmetries broken were found only in the region painted in pink. Next, as
shown below, only above the horizontal line of $|\mu +\kappa |=0.06$ one
can find stationary solutions with the broken $x\leftrightarrow -x$
symmetry, and only above the slanted nearly straight line, originating from
the origin, there exist solutions with the symmetry broken between
components $u$ and $v$. Below we derive conditions that predict these
borders, and illustrate specific cases for selected values of $\kappa $,
realizing scenarios of the spontaneous breaking of both symmetries. Note,
however, that in Fig.~(\ref{freshmap}) there is also a region where
only the solutions with preserving full symmetry or with both symmetries broken exist. In this region there are no solutions with only one symmetry broken.

\begin{figure}[th]
\centering
\includegraphics[scale=0.22]{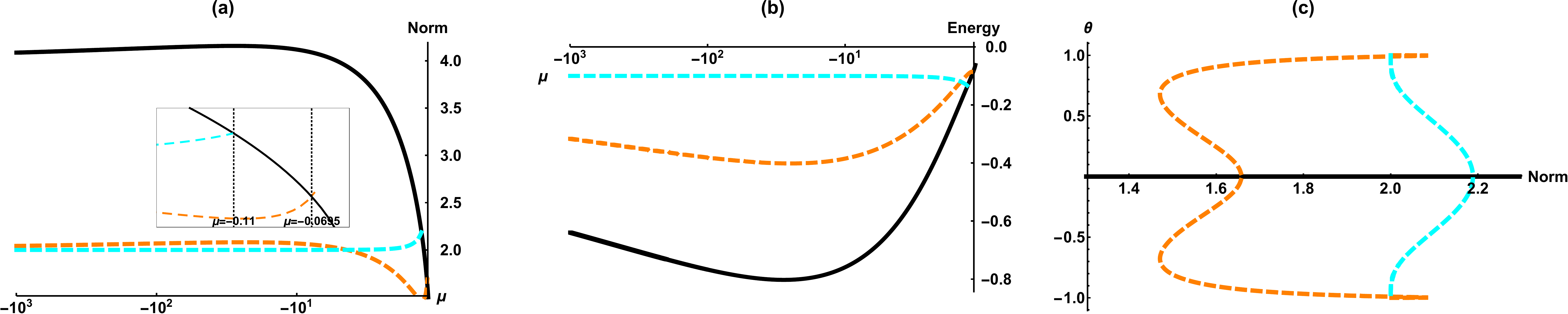}
\caption{Stationary states for a fixed value of the inter-core coupling
constant, $\protect\kappa =0.05$. Panels (a) and (b): the energy and norm of
the fully symmetric states (black), ones with broken inter-core symmetry
(orange), and states with
broken $x\leftrightarrow -x$ symmetry in each channel (blue), as functions of the chemical
potential. The inset in (a) displays details of the emergence of the
asymmetric states from the symmetric one. Panel (c) displays the
corresponding symmetry-breaking bifurcations by means of asymmetry parameters $\Theta_1$ and $\Theta_2$ (the orange
and blue curves, respectively, see Eqs. (\ref{asymmetry parameter 1})
and (\ref{asymmetry parameter 2}), plotted versus the total norm.}
\label{coupling1}
\end{figure}

\begin{figure}[th]
\centering
\includegraphics[scale=0.22]{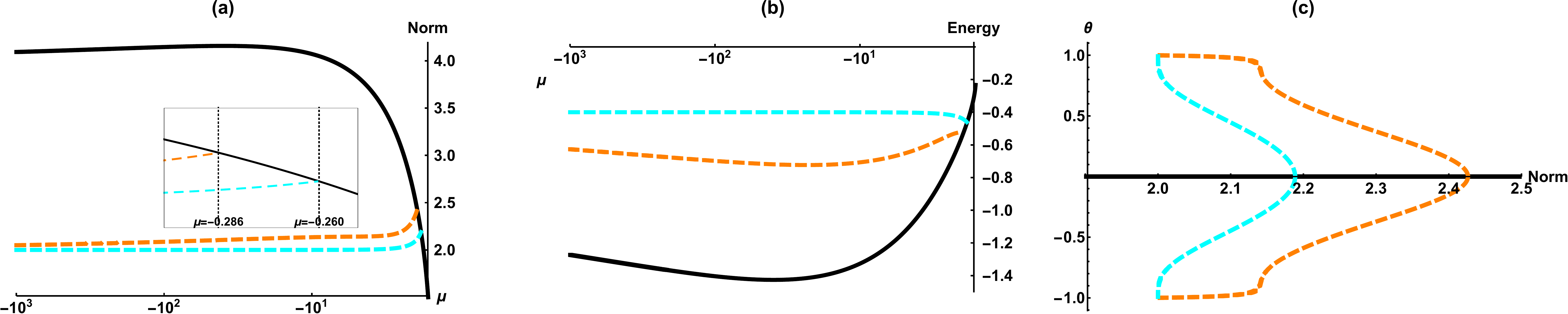}
\caption{Stationary states for fixed value of $\protect\kappa =0.2$. Panels
(a) and (b): energy and norm of the symmetric state (black), states with broken inter-core symmetry
(orange), and states with
broken $x\leftrightarrow -x$ symmetry in each channel (blue), vs. the chemical potential. The inset in (a)
displays details of the emergence of the asymmetric states from the
symmetric one. In panel (c), the corresponding symmetry-breaking
bifurcations are shown by means of asymmetry parameters $\Theta_1$ and $\Theta_2$ (the orange
and blue curves, respectively, see Eqs. (\ref{asymmetry parameter 1})
and (\ref{asymmetry parameter 2}), plotted versus the total norm.}
\label{coupling2}
\end{figure}

In the following we fixed the value of $\kappa $, and by varying the chemical potential, we sampled solutions along vertical lines in Fig.~\ref{freshmap}. Our results are outlined in Figs. \ref{coupling1} and \ref{coupling2}, for $\kappa =0.05$ and $\kappa=0.2$, respectively. In order to describe these figures we introduce asymmetry parameters $\Theta_1$ and $\Theta_2$. The asymmetry between the $u$ and $v$ components is characterized by
parameter
\begin{equation}
\Theta_1 =\frac{\int_{-\infty }^{+\infty }\left[ |u^{2}(x)|-|v^{2}(x)|\right] dx}{%
\int_{-\infty }^{+\infty }\left[ |u^{2}(x)|+|v^{2}(x)|\right] dx},
\label{asymmetry parameter 1}
\end{equation}%
and asymmetry with respect to the parity transformation, $x \longrightarrow -x$, is quantified by
\begin{equation}
\Theta_2 =\frac{\int_{-\infty }^{0 }\left[ |u^{2}(x)|+|v^{2}(x)|\right] dx - \int_{0 }^{+\infty }\left[ |u^{2}(x)|+|v^{2}(x)|\right] dx}{%
\int_{-\infty }^{+\infty }\left[ u^{2}(x)+v^{2}(x)\right] dx}.
\label{asymmetry parameter 2}
\end{equation}%

In Fig.~(\ref{coupling1}), we vary $\mu $ to the point
of crossing the slanted line in Fig.~\ref{freshmap}, at $\mu \approx -0.062$. This is the border line above which (unstable) states featuring the
asymmetry between the cores appear, only symmetric and antisymmetric states
existing below this border. These asymmetric solutions are shown by the
dashed orange line. Upon further decreasing the chemical potential we
eventually enter the region where new solutions appear, with parity symmetries
broken, which are represented by the blue dashed line in Fig. \ref%
{coupling1}. Panels (a) and (b) show the norm and energy of these states as
functions of the chemical potential, while panel (c) shows the
asymmetry parameters $\Theta_1$ and $\Theta_2$ (see Eqs. (\ref{asymmetry parameter 1})
and (\ref{asymmetry parameter 2}) as functions of the norm,
revealing that both symmetry-breaking bifurcations are of the subcritical \cite%
{bifurcations}\ type.

In Fig.~(\ref{coupling2}) the black curve of symmetric states traverses the
horizontal line into the region where (unstable) solutions, asymmetric with
respect to $x\leftrightarrow -x$, appear (the blue dashed lines), and the
second symmetry is broken at $\mu =-0.286$, to produce the states with inter-core symmetry broken (orange lines). Again, both bifurcations have the subcritical
character.

In the next two sections we investigate asymmetry solutions, to identify the
respective bifurcations. First we restrict the analysis to the class of
solutions that remain symmetric (or antisymmetric) with respect to $%
x\leftrightarrow -x$ in each core. Within this class, we consider the
possibility of the symmetry breaking between the cores. Next, we summarize
findings obtained in previous works for the single-component model, which
corresponds to the setting which maintains the symmetry or antisymmetry
between the cores ($u=\pm v$). In the case of the spatial symmetry breaking (%
$x\leftrightarrow -x$), the model predicts a subcritical type of the
spontaneous symmetry breaking. In that context, it is interesting to
investigate to what extend the structure of the model affects the transition
to the asymmetric states. To this end, we consider various forms of the
nonlinearity modulation patterns, including an additional linear potential.
Before addressing the model with the double delta function, as defined by
Eq. (\ref{delta}), we first consider the profile of $g(x)$ with a single
delta function.

\section{Solutions preserving the spatial symmetry in the cores}

\subsection{The basic model}

The first type of the symmetry breaking, which was not studied in previous
works, occurs under the assumption that the wave function remains spatially
symmetric (even) or antisymmetric (odd), with respect to $%
x\leftrightarrow -x$, in each core, while the wave functions in the two
cores may become different, $u\neq v$. As concerns coefficients in
expression (\ref{NLSEs solutions}) for the solutions, this type
of the localized states is ensured by setting $A=B$ and $C=D$. To obtain
fully symmetric states, we choose $C=D=0$; then, it is straightforward to
obtain from Eq. \ref{cubic equations} a previously known solution \cite{Dong}%
,%
\begin{equation}
A=B=\pm \sqrt{\frac{\sqrt{\mu _{+}}}{1+e^{-2\sqrt{\mu _{+}}}}}.
\label{symmetric amplitudes}
\end{equation}%
Likewise, for antisymmetric states we assume $A=B=0$ to obtain

\begin{equation}
C=D=\pm\sqrt{\frac{\sqrt{\mu_{-}}}{1+e^{-2\sqrt{\mu_{-}}}}}
\label{antisymmetric amplitudes}
\end{equation}

Searching for states asymmetric between the parallel-coupled cores, we
assume that all coefficients $A=B$ and $C=D$ are different from zero. In
this case, solutions of Eq. (\ref{cubic equations}) are

\begin{eqnarray}
A &=&B=\pm \sqrt{\frac{3\sqrt{\mu _{-}}\left( 1+e^{-2\sqrt{\mu _{+}}}\right)
-\sqrt{\mu _{+}}\left( 1+e^{-2\sqrt{\mu _{-}}}\right) }{8\left( 1+e^{-2\sqrt{%
\mu _{-}}}\right) \left( 1+e^{-2\sqrt{\mu _{+}}}\right) }},  \notag \\
C &=&D=\pm \sqrt{\frac{3\sqrt{\mu _{+}}\left( 1+e^{-2\sqrt{\mu _{-}}}\right)
-\sqrt{\mu _{-}}\left( 1+e^{-2\sqrt{\mu _{+}}}\right) }{8\left( 1+e^{-2\sqrt{%
\mu _{-}}}\right) \left( 1+e^{-2\sqrt{\mu _{+}}}\right) }}.
\label{amplitudes for symmetry of each mode separately}
\end{eqnarray}%
To find the symmetry-breaking bifurcation point, we inspect the second
expression in Eq. \ref{amplitudes for symmetry of each mode separately},
equating the numerator under the square root to zero. This leads to

\begin{equation}
3\sqrt{\frac{|\mu _{\mathrm{bif}}+\kappa |}{|\mu _{\mathrm{bif}}-\kappa |}} = %
\frac{1+e^{-2\sqrt{2|\mu _{\mathrm{bif}}+\kappa |}}}{1+e^{-2\sqrt{2|\mu _{%
\mathrm{bif}}-\kappa |}}},  \label{inequality for asymmetric states}
\end{equation}%
which is the main result of this section.
It corresponds to the almost straight slanted border line in Fig. (\ref%
{freshmap}). We stress that, in the present model, all states besides totally symmetric ones are unstable.
In particular, the states that exist above the line defined by Eq.~(\ref{inequality for asymmetric states}), which are asymmetric with respect to the two cores ($u \neq v$), are unstable. Actually, this
is the property of this particular model. In Section 5 we introduce a two-dimensional model, in
which broken-symmetry solutions are stable.

We supplement this section by the following subsection, revealing the type of the symmetry-breaking
bifurcation that one may expect
dealing with models consisting of two identical cores and various
nonlinearity-modulation patterns or linear potentials.

\subsection{Review of related models}

We first consider the case of the simplified spatial modulation with
\begin{equation}
g\left( x\right) = - \delta \left( x\right) ,  \label{single}
\end{equation}%
representing one singularity instead of two. In this case, exact solution
\eqref{NLSEs solutions} is replaced by

\begin{eqnarray}
u &=&\left\{
\begin{array}{c}
Ae^{\sqrt{\mu _{+}}x}+Ce^{\sqrt{\mu _{-}}x},~\mathrm{at}~~x<0, \\
Ae^{-\sqrt{\mu _{+}}x}+Ce^{-\sqrt{\mu _{-}}x},~\mathrm{at}~~x>0,%
\end{array}%
\right.  \notag \\
v &=&\left\{
\begin{array}{c}
Ae^{\sqrt{\mu _{+}}x}-Ce^{\sqrt{\mu _{-}}x},~\mathrm{at}~~x<0, \\
Ae^{-\sqrt{\mu _{+}}x}-Ce^{-\sqrt{\mu _{-}}x},~\mathrm{at}~~x>0.%
\end{array}%
\right.  \label{reduced problem solutions}
\end{eqnarray}%
The conditions for the derivative jump at $x=0$ are $\Delta \left( u^{\prime
}\right) |_{x=0}=-2\left( u|_{x=0}\right) ^{3}$ and $\Delta \left( v^{\prime
}\right) |_{x=0}=-2\left( v|_{x=0}\right) ^{3}$, cf. Eq. (\ref{Delta}). This
leads to the system of coupled cubic equations

\begin{eqnarray}
\sqrt{\mu _{+}}A+\sqrt{\mu _{-}}C &=&\left( A+C\right) ^{3},  \notag \\
\sqrt{\mu _{+}}A-\sqrt{\mu _{-}}C &=&\left( A-C\right) ^{3},
\end{eqnarray}%
cf. Eq. (\ref{cubic equations}), which can be easily solved. For the
asymmetric solution, with $A\neq 0$ and $C\neq 0$, the amplitudes are
\begin{eqnarray}
A &=&\pm \frac{1}{2}\sqrt{\frac{3}{2}\sqrt{\mu _{-}}-\frac{1}{2}\sqrt{\mu
_{+}}},  \notag \\
C &=&\pm \frac{1}{2}\sqrt{\frac{3}{2}\sqrt{\mu _{+}}-\frac{1}{2}\sqrt{\mu
_{-}}}.  \label{simplified amplitudes}
\end{eqnarray}%
The symmetry-breaking bifurcation point is derived by setting $C=0$ in Eq. (%
\ref{simplified amplitudes}), which yields

\begin{equation}
\mu _{\mathrm{bif}}=-\frac{5}{4}\kappa .  \label{bifurcation inequality}
\end{equation}%
Note that, when the inter-core coupling is negligible, i.e. $\kappa =0$, the
asymmetric states have $\mu <0$, hence they exist everywhere in the
trapped-mode regime. On the other hand, the symmetry breaking does not take
place for $\kappa \rightarrow \infty $, the solution keeping the form of $%
u=v $.

\begin{figure}[th]
\centering
\subfigure[]{\includegraphics[scale=0.5]{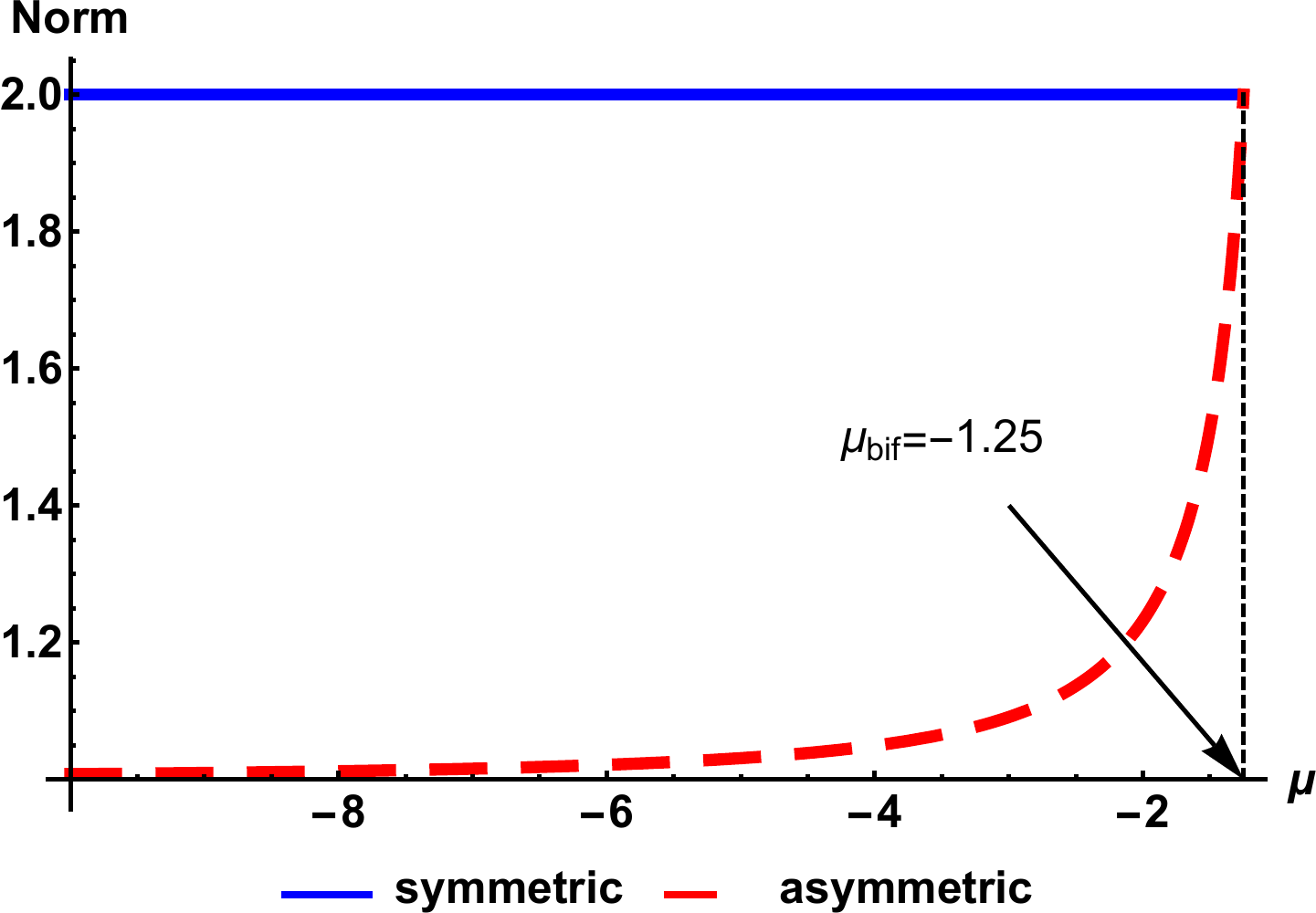}} %
\subfigure[]{\includegraphics[scale=0.5]{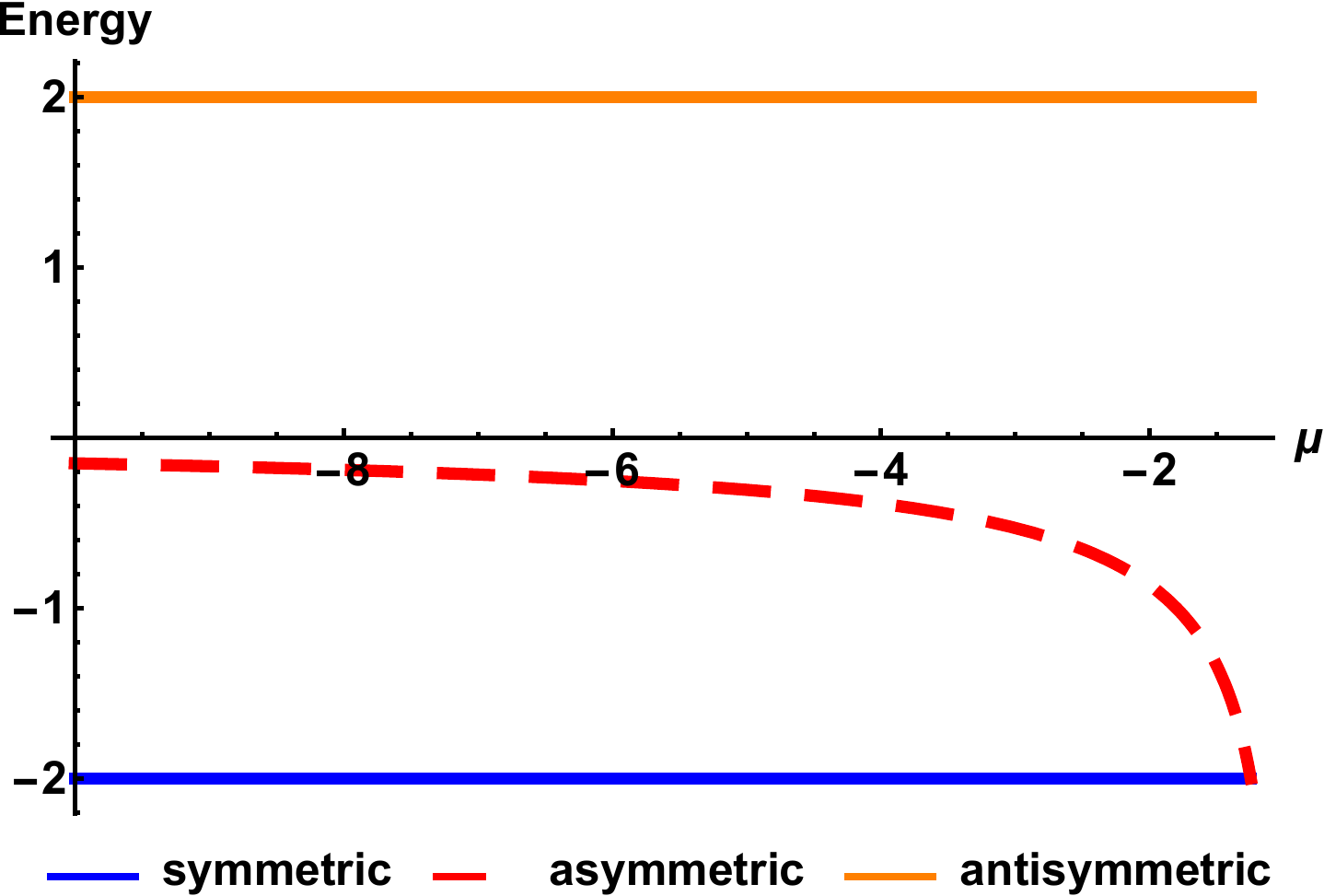}}
\caption{Norm (a) and energy (b) of the symmetric, antisymmetric and
asymmetric states as a function of the chemical potential in the model with
the nonlinearity modulation in the form of the single delta-function, as
defined by Eq. (\protect\ref{single}). Solid and dashed lines denote stable
and unstable states, respectively.}
\label{1delta_enorm}
\end{figure}

Amplitudes of the symmetric and antisymmetric states ($u=v$ and $u=-v$) can
be easily found too. In particular, amplitudes of the symmetric state amount
to $A^{2}=\sqrt{\mu _{+}/2}$ and $C^{2}=\sqrt{\mu _{-}/2}$, and norm (\ref{N}%
) of symmetric and antisymmetric states is $N_{\mathrm{symm}}=N_{\mathrm{%
antisymm}}=2$. The norm of the asymmetric state is
\begin{equation}
N_{\mathrm{asymm}}=-\frac{3\mu +\sqrt{\mu ^{2}-\kappa ^{2}}}{2\sqrt{\mu
^{2}-\kappa ^{2}}},  \label{asymmetric simplified real norm}
\end{equation}%
with limit value $N_{\mathrm{asymm}}(\mu \rightarrow -\infty )=1$. The norms
of the symmetric and asymmetric modes are plotted, as functions of $\mu $,
in Fig. \ref{1delta_enorm}(a).

\begin{figure}[th]
\centering
\subfigure[]{\includegraphics[scale=0.5]{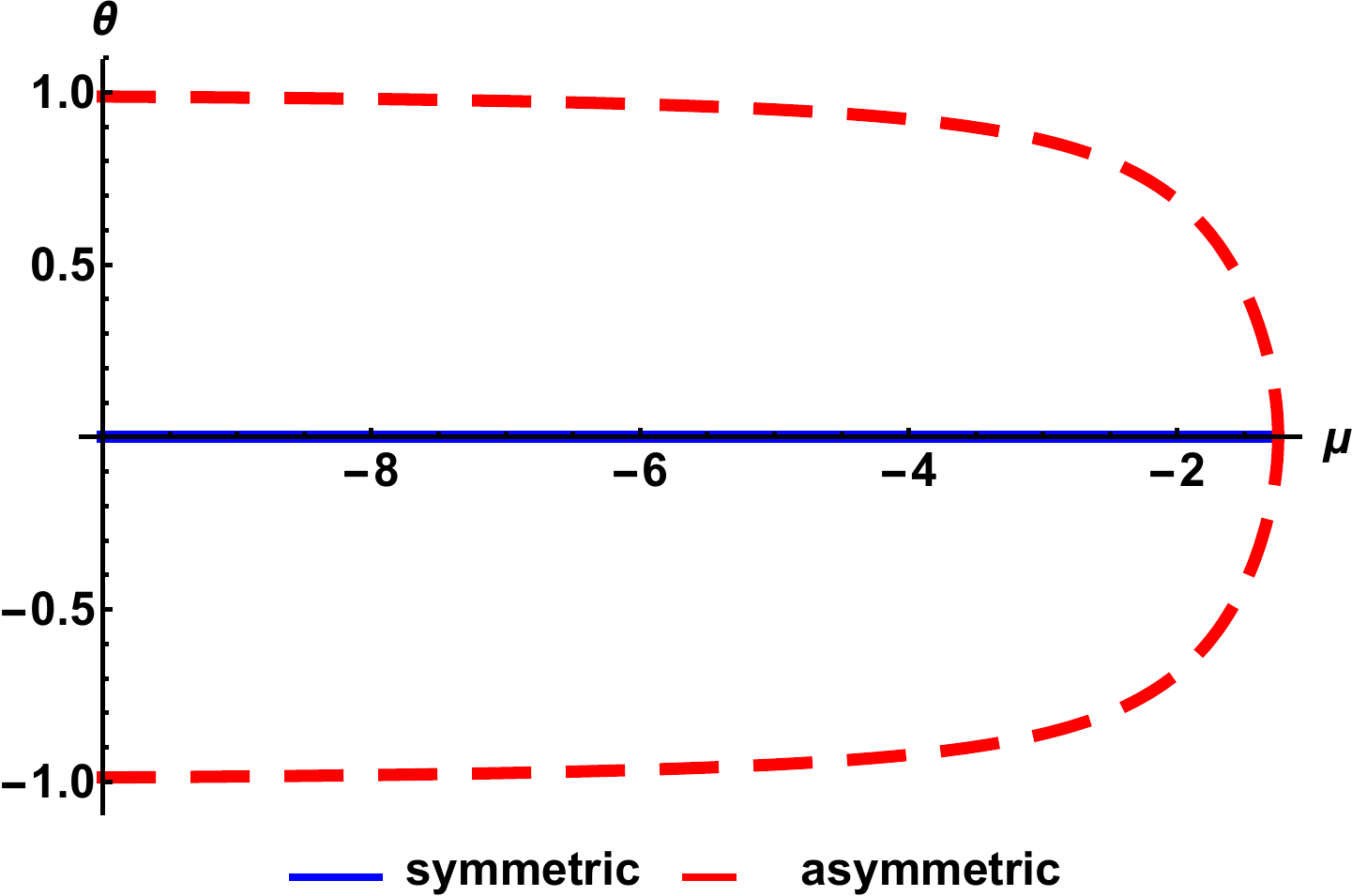}} %
\subfigure[]{\includegraphics[scale=0.5]{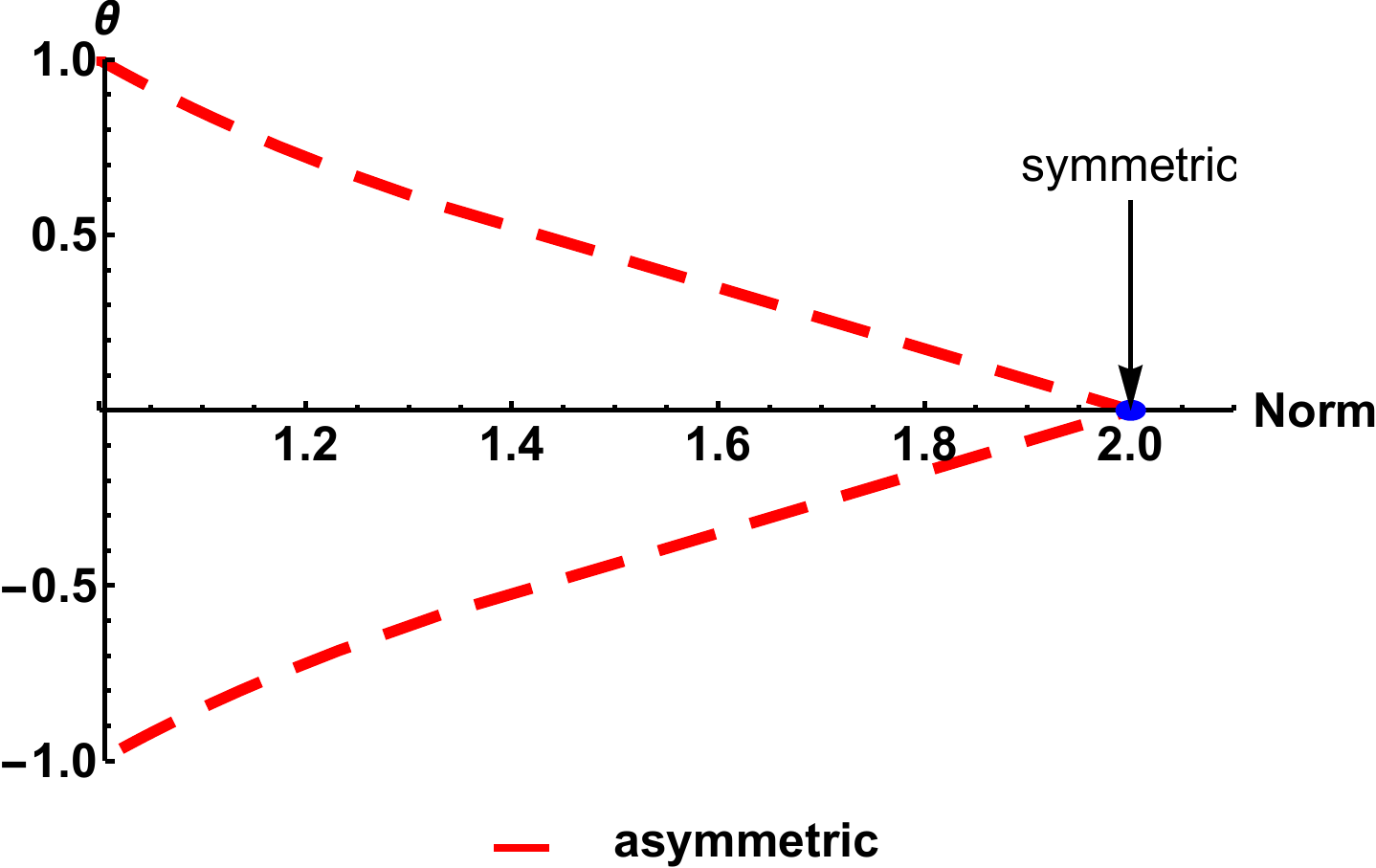}}
\caption{Asymmetry parameter $\Theta $, defined as per Eq.~(\protect\ref%
{asymmetry for single delta}), as a function of the chemical potential (a)
and total norm (b). Solid and dashed lines correspond to stable and unstable
states, respectively.}
\label{1delta_asymetry}
\end{figure}

The value of the energy (Hamiltonian), defined in Eq. (\ref{H}), can also be
calculated for all the states under the consideration, a general expression being

\begin{equation}
E=\sqrt{\mu _{+}}|A|^{2}+\sqrt{\mu _{-}}|C|^{2}-|A|^{4}-A^{2}\left( C^{\ast
}\right) ^{2}  \notag  \label{hamilton} \\
-4|A|^{2}|C|^{2}-\left( A^{\ast }\right) ^{2}C^{2}-|C|^{4}-\frac{\kappa
|A|^{2}}{\sqrt{\mu _{+}}}+\frac{\kappa |C|^{2}}{\sqrt{\mu _{-}}}.
\end{equation}%
In the special cases of the symmetric, antisymmetric and asymmetric states
the corresponding energies are $E_{\mathrm{symm}}=-\kappa $ and $E_{\mathrm{%
antisymm}}=\kappa $, and

\begin{equation}
E_{\mathrm{asymm}}=-\frac{3}{8}\kappa \left[ \sqrt{\frac{\mu _{-}}{\mu _{+}}}%
-\sqrt{\frac{\mu _{+}}{\mu _{-}}}\right] .
\label{asymmetric energy for single delta}
\end{equation}%
At the bifurcation point (see Fig. \ref{1delta_enorm}(b)), $\mu =-\left(
5/4\right) \kappa $, expression \eqref{asymmetric energy for single delta}
coincides with $E_{\mathrm{symm}}=-\kappa $.

Finally, asymmetry coefficient (\ref{asymmetry parameter 1}), which measures
the norm imbalance between the cores, is calculated for solution (\ref%
{simplified amplitudes}) as

\begin{equation}
\Theta =\frac{4\sqrt{5\sqrt{\mu ^{2}-\kappa ^{2}}+3\mu }\cdot \sqrt{\mu
^{2}-\kappa ^{2}}}{\left( -3\mu -\sqrt{\mu ^{2}-\kappa ^{2}}\right) \left(
\sqrt{-2(\mu +\kappa )}+\sqrt{-2(\mu -\kappa )}\right) }.
\label{asymmetry for single delta}
\end{equation}%
It is plotted versus the chemical potential and norm in Fig. \ref%
{1delta_asymetry}. Note that $\Theta \rightarrow 1$ in the limit of ${\mu
\rightarrow -\infty }.$ All the asymmetric states are unstable, in the model
based on the single-delta modulation format (\ref{single}).

Numerical investigation was also performed for similar models with the
single delta function replaced by a finite-width Gaussian. In this case, the
symmetry-breaking bifurcation keeps the subcritical form even in the case
of a broad Gaussian, cf. Fig. \ref{1delta_asymetry}(b). This finding is in
agreement with the fact that, in various models of dual-core couplers with
the uniform self-focusing nonlinearity, symmetry-breaking bifurcations for
solitons are subcritical too \cite{Laval}-\cite{Skinner}, \cite{bifurcations,Hung}.
Additionally, we have investigated the model of the coupled cores with the uniform self-focusing nonlinearity, combined with a local defect represented by a linear-potential well (not shown here in detail). In this case, we have
found the subcritical transition in shallow wells, regardless of their
width, which carry over into a supercritical transition when the well's
depth exceeds a certain critical value.

\section{The case of the symmetry maintained between the cores}

Here we present a short summary of results obtained when we impose the
condition of the symmetry between the cores, i.e. $u=\pm v$, comparing this
case to the study presented in Ref. \cite{Dong}, where the spontaneous
symmetry breaking in was due to the action of the nonlinear potential with
two symmetric minima. When the respective modulation pattern was taken as
per Eq. (\ref{delta}), a fully analytical solution was obtained for
symmetric, antisymmetric, and asymmetric states, the respective
symmetry-breaking bifurcation was \textit{ultimately subcritical}, with the
branches of asymmetric localized states never turning in the forward
direction.

In terms of Eqs. (\ref{cubic equations}), the cases of $u=v$ and $u=-v$
correspond, respectively, to setting $C=D=0$ and $A=B=0$. In the former
(symmetric) case the remaining equations for amplitudes $A$ and $B$ rare
\begin{eqnarray}
\sqrt{\mu _{+}}e^{2\sqrt{\mu _{+}}}\left( e^{2\sqrt{\mu _{+}}}B-A\right)
&=&\left( e^{4\sqrt{\mu _{+}}}-1\right) B^{3},  \notag \\
\sqrt{\mu _{+}}e^{2\sqrt{\mu _{+}}}\left( e^{2\sqrt{\mu _{+}}}A-B\right)
&=&\left( e^{4\sqrt{\mu _{+}}}-1\right) A^{3}.  \label{Boris case}
\end{eqnarray}%
To treat the antisymmetric, one may replace $A\rightarrow C$, $B\rightarrow
D $ and $\mu _{+}\rightarrow \mu _{-}$. The respective solutions are

\begin{equation}
A=B\equiv A_{\mathrm{sym}}=\pm \sqrt{\frac{\sqrt{\mu _{+}}}{1+e^{-2\sqrt{\mu
_{+}}}}},
\end{equation}%
and an antisymmetric form (i.e. antisymmetry in each channel)

\begin{equation}
A=-B\equiv A_{\mathrm{antisymm}}=\pm \sqrt{\frac{\sqrt{\mu _{+}}}{1-e^{-2%
\sqrt{\mu _{+}}}}}.
\end{equation}%
The norm of these solutions is
\begin{equation}
N_{\mathrm{symm},\mathrm{antisymm}}=\frac{1}{1\pm e^{-2\sqrt{\mu _{+}}}}+%
\frac{1-e^{-4\sqrt{\mu _{+}}}\pm 4\sqrt{\mu _{+}}e^{-2\sqrt{\mu _{+}/2}}}{%
\left( 1\pm e^{-2\sqrt{\mu _{+}/2}}\right) },
\end{equation}%
where $+$ and $-$ correspond to the symmetric and antisymmetric ones,
respectively. Both solutions are characterized by the same asymptotic limit,
$N_{\mu \rightarrow -\infty }\rightarrow 2$. The chemical potential at the
bifurcation point is evaluated, for the symmetric solution, (in agreement
with \cite{Dong}) as
\begin{equation}
\tilde{\mu}_{\mathrm{bif}}=-\left( \ln 2\right) ^{2}/8\approx 0.06,
\label{bif}
\end{equation}
while the antisymmetric one undergoes no bifurcation. Likewise, the
amplitudes for asymmetric states can be found analytically,
\begin{equation}
\left\{ A,B\right\} _{\mathrm{asymm}}=\frac{\sqrt[4]{\mu _{+}/2}\left( \sqrt{%
1+2e^{-2\sqrt{\mu _{+}}}}\pm \sqrt{1-2e^{-2\sqrt{\mu _{+}}}}\right) }{2^{3/4}%
\sqrt{1-e^{-4\sqrt{\mu _{+}}}}},
\end{equation}%
and the bifurcation point derived from this solution is identical to one
given by Eq. (\ref{bif}).

In the case when symmetry $u=\pm v$ is imposed, we can compare our results
with the more realistic model of with the Gaussian modulation format \cite%
{Dong},

\begin{equation}
g\left( x\right) =\frac{-1}{a\sqrt{\pi }}\left[ \exp \left( -\frac{\left(
x+1\right) ^{2}}{a^{2}}\right) +\exp \left( -\frac{\left( x-1\right) ^{2}}{%
a^{2}}\right) \right] ,  \label{g(x)}
\end{equation}%
where the modulated nonlinearity coefficient is subject to the normalization
condition $\int_{-\infty }^{+\infty }g\left( x\right) dx=2.$ It is found
that, at a certain value of width $a$ of the Gaussian pattern (\ref{g(x)}),
a transition to the supercritical type of the symmetry-breaking bifurcation
takes place. On the other hand, if we consider systems with uniform
self-focusing nonlinearity in the presence of a linear potential
representing a double potential well \cite{DWP1,DWP2,Marek}, spontaneous
symmetry breaking occurs too, but it always has the supercritical character.

\section{Double symmetry breaking in two dimensions}

\begin{figure}[th]
\centering
\includegraphics[scale=0.4]{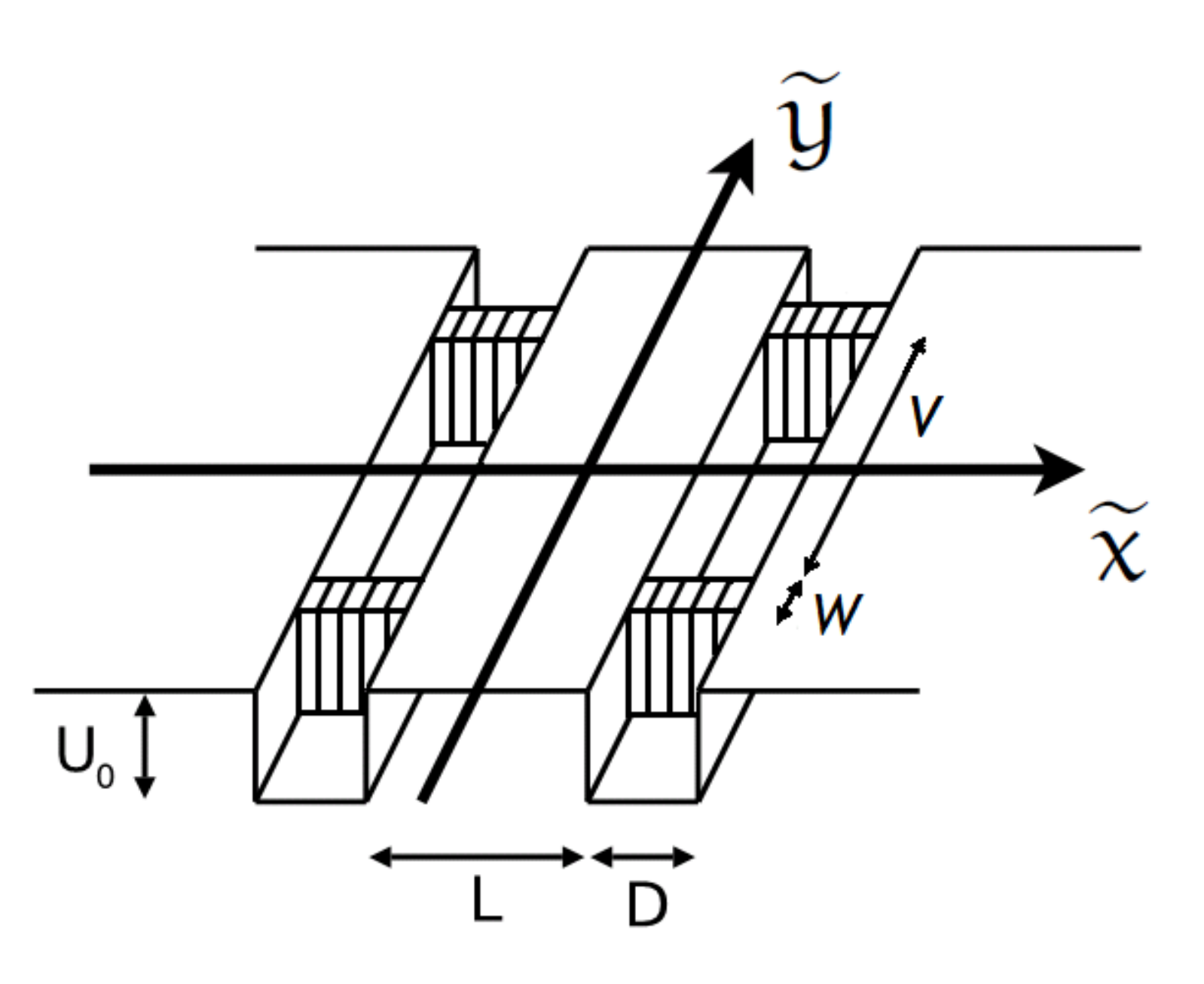}
\caption{The combination of the linear-potential double-trough trapping
acting along the $x$ direction, and the localized nonlinearity pattern shown
by the striped structure. We address double symmetry breaking of
two-dimensional solitons in this setting.}
\label{potential2d}
\end{figure}

In the previous sections we considered solutions of coupled one-dimensional
systems with the nonlinearity concentrated at two narrow spots, finding
stationary states with one of the symmetries broken (ether between the
parallel-coupled cores, or between the two nonlinear spots in the cores), or
even the situation when both underlying symmetries are broken in the
stationary state. In most cases, however, and in all cases when the
nonlinearity modulations is based on the pair of delta-functions (Eq. (\ref%
{delta})), states with broken symmetry were unstable. One can argue that the
model with the ideal delta-functions is degenerate, which is reflected in
the character of the stationary states. In this section, we do not consider
bifurcations and spontaneous symmetry breaking for settings with the
Gaussian nonlinearity-modulation profiles, replacing the two
delta-functions. Instead of that, in this section we focus on a
two-dimensional model with localized nonlinearity, showing that in this case
states with broken symmetries may be stable. The setting that we address is
based on the following GPE,
\begin{equation}
i\frac{\partial \psi }{\partial t}=-\frac{1}{2}\left( \frac{\partial
^{2}\psi }{\partial x^{2}}+\frac{\partial ^{2}\psi }{\partial y^{2}}\right)
+g\left( x,y\right) |\psi |^{2}\psi +U(x)\psi .  \label{2D model}
\end{equation}%
As displayed in Fig. \ref{potential2d}, potential $U(x)$ creates the
double-well (double-trough) trapping potential acting in the $x$ direction,
while nonlinearity coefficient $g\left( x,y\right) $ is concentrated at four
localized spots (striped in the figure). This model may be considered as a
two-dimensional extension of the one-dimensional model introduced above.

\begin{figure}[th]
\centering
\subfigure{\includegraphics[scale=0.5]{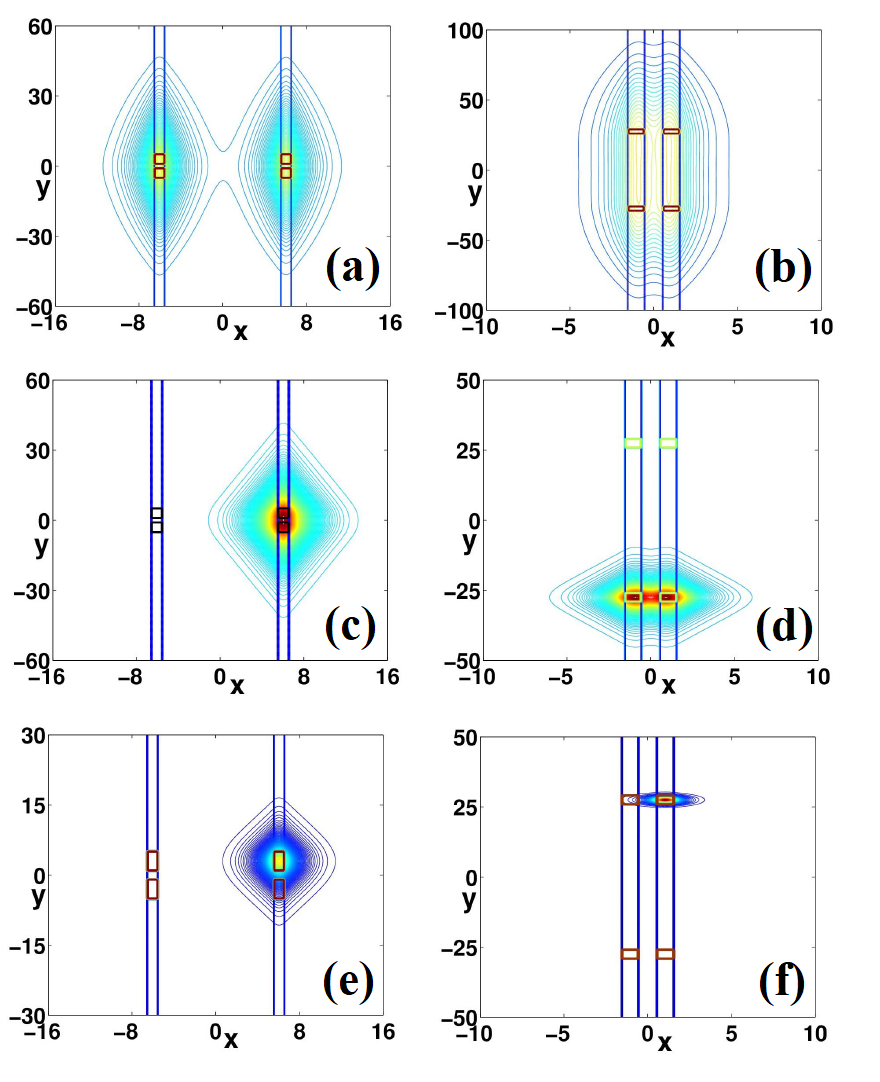}}
\caption{Contour plots of the local density illustrating stable
two-dimensional self-trapped states. Panels (a) and (b): fully symmetric
ones. (c) and (d): States with broken symmetry between the two cores, and
between the two nonlinear spots, respectively. States featuring the breaking
of both symmetries are displayed in panels (e) and (f).}
\label{hung1}
\end{figure}

We used the imaginary-time method \cite{IT} to find ground states of the system. Their stability was then confirmed by simulating the real time evolution (using the split-step Fourier method). We have observed  four types of stable solitons, which are shown in Fig.~\ref{hung1}. The figure confirms that, depending on the value of the norm, stable states with all possible combinations of the asymmetry can be produced.
A general conclusion, suggested by the systematic simulations, is that, in the model with weaker coupling between the cores, it is easier to find stable states with the broken symmetry between the cores. With the increase of the norm, states with both symmetries broken are found too. On the other hand, the symmetry between the nonlinearity-concentration spots in both cores in broken first in the setting with the nonlinearity spots set closer to each other. In particular, values of parameters in Fig.~\ref{hung1} are:\\
in Figs. \ref{hung1} (a), (c) and (e), $L=11$, $D=1$, $W=4$, $V=2$,
and the norm in the increasing order, $N=0.6$, $N=1.5$, and $N=2.5$,
respectively. Similarly, in Fig. \ref{hung1} (b), (d) and (f), the
parameters are $L=1$, $D=1$, $W=3$, $V=52$, and norms are,
respectively, $N=1$, $4$, and $5.1$. The depth of linear potential $U_0$ and the (constant) value of the nonlinear coupling can be rescaled. The strength of the interaction is now contained in the norm of the wavefunction.

In the real-time simulations, we added small perturbations to check if the input states
maintain their initial shapes after long evolution. All of
the above states still survive, at least, up to
$t=1000$.

Our two-dimensional model surely deserves further investigation, including the identification of bifurcation diagrams, regions of stability of various types of solutions, and studies of dynamical behavior, including
collisions between solitons. Detailed results addressing these issues will be reported elsewhere.

\section{Conclusion}

The objective of this work is to investigate the phenomenology of
spontaneous symmetry breaking in the class of dual-core systems featuring
the additional spatial symmetry in each core, induced by the
nonlinearity-modulation pattern in the form of two tightly localized spots,
which may be approximated, in some cases, by ideal delta-functions. In this
case, solutions for stationary one-dimensional states can be obtained in the
implicit analytical form, represented by coupled cubic equations for
constituent amplitudes. The analysis makes it possible to construct modes
with broken inter-core and intra-core symmetries, as well as ones featuring
the \textit{double breaking} of both symmetries. While all such asymmetric
modes turn out to be unstable, the consideration of the most interesting
two-dimensional version of the system, with a small transverse thickness of
each core, and a small finite size of the nonlinearity-localization spots,
reveals \emph{stable} asymmetric states of all the types, including ones
realizing the double symmetry breaking.

As an extension of the analysis, it may be interesting to construct
stationary patterns in which the complex wave function, peaked at the
nonlinearity-localization spots, features a phase structure which
corresponds to \textit{trapped vorticity}, following the pattern of Ref.
\cite{Kevrekidis}. The existence and stability of such vortex modes, with
unbroken or broken symmetries, is a challenging issue.

\vspace{6pt}

\acknowledgments{The work was supported by the Polish National
Science Centre 2016/22/M/ST2/00261 (A. Ramaniuk and M. Trippenbach.)
K.B.Z. acknowledges support from the National Science Center of
Poland through Project FUGA No. 2016/20/S/ST2/00366. N.V.H. was
supported by Vietnam National Foundation for Science and Technology
Development (NAFOSTED) under grant number 103.01-2017.55.}

\end{document}